# Attosecond magnetization dynamics in non-magnetic materials driven by intense femtosecond lasers

Ofer Neufeld[1,*], Nicolas Tancogne-Dejean[1], Umberto De Giovannini[1,2], Hannes Hübener[1], and Angel Rubio[1,3,*]

[1]Max Planck Institute for the Structure and Dynamics of Matter and Center for Free-Electron Laser Science, Hamburg, Germany, 22761.
[2]Università degli Studi di Palermo, Dipartimento di Fisica e Chimica—Emilio Segrè, Palermo I-90123, Italy.
[3]Center for Computational Quantum Physics (CCQ), The Flatiron Institute, New York, NY, USA, 10010.
*Corresponding author E-mails: ofer.neufeld@gmail.com, angel.rubio@mpsd.mpg.de

Irradiating solids with ultrashort laser pulses is known to initiate femtosecond timescale magnetization dynamics. However, sub-femtosecond spin dynamics have not yet been observed or predicted. Here, we explore ultrafast light-driven spin dynamics in a highly non-resonant strong-field regime. Through state-of-the-art *ab-initio* calculations, we predict that a non-magnetic material can be transiently transformed into a magnetic one *via* dynamical extremely nonlinear spin-flipping processes, which occur on attosecond timescales and are mediated by a combination of multi-photon and spin-orbit interactions. These are non-perturbative non-resonant analogues to the inverse Faraday effect that build up from cycle-to-cycle as electrons gain angular momentum. Remarkably, we show that even for linearly polarized driving, where one does not intuitively expect any magnetic response, the magnetization transiently oscillates as the system interacts with light. This oscillating response is enabled by transverse anomalous light-driven currents in the solid, and typically occurs on timescales of ~500 attoseconds. We further demonstrate that the speed of magnetization can be controlled by tuning the laser wavelength and intensity. An experimental set-up capable of measuring these dynamics through pump-probe transient absorption spectroscopy is outlined and simulated. Our results pave the way for new regimes of ultrafast manipulation of magnetism.

## I. INTROUCTION

Magnetism is one of the most fundamental physical phenomena in nature. It arises from internal spin degrees of freedom of quantum particles [1,2], and can create complex spin textures such as skyrmions [3–6] or other magnetic and topological ordered phases [7–9]. Over the past two decades immense efforts have been devoted towards the study of ultrafast magnetism, i.e. the manipulation of materials' magnetic structures on femtosecond timescales with ultrashort laser pulses [10–15]. However, despite many years of research numerous open questions remain regarding the mechanisms and pathways that control ultrafast magnetization dynamics [10–19]. Part of the complexity arises because spin dynamics are often entangled with other processes and interactions (because spin is carried by charged particles that also interact with each other and with other particles), and can also evolve over several orders of magnitudes of timescales ranging from femtoseconds to nanoseconds. For instance, only recently the mechanism through which angular momentum transfers from the spin order to the lattice during demagnetization was uncovered [20].

In a typical femto-magnetism experiment, an intense laser pulse is irradiated onto a magnetic material such as a ferromagnet, which initiates demagnetization dynamics, spin transfer dynamics, or spin switching [21–30]. The ultrafast dynamics are subsequently tracked through time-resolved pump-probe spectroscopy [31–34]. Recently it was shown that magnetization can even be transiently induced in non-magnetic materials with resonant circularly polarized light through the inverse Faraday effect [35], which is a perturbative nonlinear optical effect [36,37], or through slower spin-phonon couplings [38]. However, to our knowledge all experiments and theoretical works thus far have never observed or predicted the following: (i) An effect whereby a non-magnetic material is illuminated by an intense non-resonant laser pulse that initiates a few femtosecond turn-on of magnetization. (ii) Magnetization dynamics driven in non-magnetic materials by linearly polarized light. (iii) Magnetization dynamics that occurs on sub-femtosecond timescales. All of these effects could pave the way to new regimes in ultrafast and non-equilibrium magnetism, e.g. allowing extremely fast manipulation of magnetic orders even in materials that have a non-magnetic ground-state.

In parallel to advancements in femto-magnetism, strong-field physics in solids has developed as a novel approach for controlling electron motion on sub-femtosecond timescales [39–45]. Strong-field interactions in solids enabled tailoring valley pseudo-spin occupations [46,47], controlling material topological properties [46], steering Dirac electrons [48] and more [49–52]. This regime provides an ideal setting for exploring possibilities of attosecond magnetism, because it gives natural access to attosecond electron motion (whereby electrons act as spin carriers). To capitalize this, a strong spin-orbit interaction could allow converting electronic angular momenta into magnetism (because light does not directly couple to spin degrees of freedom). Extremely nonlinear light-matter interactions such as high harmonic generation (HHG) have been explored in some material systems with strong spin-orbit interactions (e.g. in $BiSbTeSe_2$ [53], $Bi_2Te_3$ [54], $Bi_2Se_3$ [55], $Ca_2RuO_4$ [56], $Na_3Bi$ [57]), but the induced magnetization was not investigated.

Here we report on femto-magnetic phenomena that are driven in non-magnetic materials by intense ultrashort laser pulses in the strong-field and highly nonlinear regime of light-matter interactions. We demonstrate with state-of-the-art time-dependent spin density functional theory calculations that strong magnetization of ~0.1$\mu_B$ (where $\mu_B$ is a Bohr magneton) can be turned-on extremely fast when driven by non-resonant circularly polarized light, within ~16 femtoseconds. This transient magnetic state is expected to live for several tens of femtoseconds before it is destroyed by scattering and dephasing. We thoroughly analyze this effect and show that the magnetization arises from highly nonlinear multi-photon processes, which together with spin-orbit interactions, allow for attosecond spin polarization to build up over time. We also study systems irradiated by linearly polarized pulses, whereby one intuitively does not expect a magnetic response (because there is no angular momentum in the driving pulses). Remarkably, we show that even linearly polarized pulses, when sufficiently intense, can induce magnetization dynamics, owing to an interplay of electronic currents driven along the laser polarization axis and transverse anomalous currents that arise in some material systems (e.g. from a nonzero Berry curvature or other structural asymmetry). Together, these currents give rise to a sub-cycle electronic orbital angular momentum that is converted to transient attosecond magnetism. The spin expectation values can flip sign from a maximum of +0.01$\mu_B$ to a minimum of -0.01$\mu_B$ in just ~411 attoseconds. Strikingly, the speed of magnetism can be tuned by changing the laser parameters. We outline and simulate a circular dichroism attosecond transient absorption spectroscopy set-up that is capable of measuring these unique phenomena.

## II. METHODOLOGY

We begin by outlining our methodological approach. To model light-induced magnetization dynamics, we employ *ab-initio* calculations based on time-dependent spin density functional theory (TDSDFT) in the Kohn-Sham (KS) formulation [58]. The system's ground-state is directly obtained within spin-polarized DFT, and is then propagated in real-time with the following equations of motion (we use atomic units unless stated otherwise):

$$i\partial_t |\psi_{n,\mathbf{k}}^{KS}(t)\rangle = \left( \frac{1}{2}\left(-i\nabla + \frac{\mathbf{A}(t)}{c}\right)^2 \sigma_0 + v_{KS}(t) \right) |\psi_{n,\mathbf{k}}^{KS}(t)\rangle \qquad (1)$$

where $|\psi_{n,\mathbf{k}}^{KS}(t)\rangle$ is the KS-Bloch state at *k*-point **k** and band index *n*, which is a Pauli spinor:

$$|\psi_{n,\mathbf{k}}^{KS}(t)\rangle = \begin{bmatrix} |\varphi_{n,\mathbf{k},\uparrow}^{KS}(t)\rangle \\ |\varphi_{n,\mathbf{k},\downarrow}^{KS}(t)\rangle \end{bmatrix} \qquad (2)$$

with $|\varphi_{n,\mathbf{k},\alpha}^{KS}(t)\rangle$ the spin-up/spin-down part of the KS states with spin index $\alpha$. $\sigma_0$ in Eq. (1) is a 2×2 identity matrix, and $\mathbf{A}(t)$ is the vector potential of the impinging laser pulse within the dipole approximation such that $-\partial_t \mathbf{A}(t) = c\mathbf{E}(t)$, and *c* is the speed of light (in atomic units $c\approx137.036$). $v_{KS}(t)$ is the time-dependent KS potential given by:

$$v_{KS}(t) = \int d^3r' \frac{n(\mathbf{r}',t)}{|\mathbf{r}-\mathbf{r}'|}\sigma_0 + v_{XC}[\rho(\mathbf{r},t)] + V_{ion} \qquad (3)$$

where the first term in Eq. (3) is the classical Hartree term – an electrostatic mean-field interaction between electrons, where $n(\mathbf{r},t)=\sum_{n,\mathbf{k},\alpha} w_\mathbf{k}|\langle\mathbf{r}|\varphi_{n,\mathbf{k},\alpha}^{KS}(t)\rangle|^2$ is the time-dependent electron density, with $w_\mathbf{k}$ the $k$-point weights and the sum running over occupied bands. The second term in brackets, $v_{XC}$, is the exchange-correlation (XC) potential that in the local spin density approximation is a functional of the spin density matrix:

$$\rho(\mathbf{r},t) = \frac{1}{2}n(\mathbf{r},t)\sigma_0 + \frac{1}{2}\mathbf{m}(\mathbf{r},t)\cdot\boldsymbol{\sigma} \qquad (4)$$

where $\mathbf{m}(\mathbf{r},t)$ is the time-dependent magnetization vector:

$$\mathbf{m}(\mathbf{r},t) = \sum_{n,\mathbf{k}} w_\mathbf{k} \langle\psi_{n,\mathbf{k}}^{KS}(t)|\mathbf{r}\rangle \boldsymbol{\sigma} \langle\mathbf{r}|\psi_{n,\mathbf{k}}^{KS}(t)\rangle \qquad (5)$$

$V_{ion}$ in Eq. (3) represents the interactions of electrons with the lattice ions and core electrons. To reduce numerical costs we employ the frozen core approximation, and the bare Coulomb interaction between electrons and ions is replaced with a fully-relativistic nonlocal norm-conserving pseudopotential [59]. This term also includes the full *ab-initio* description of relativistic corrections to the Hamiltonian, including the mass term and the Darwin term. Most importantly, it incorporates a spin-orbit coupling term that is proportional to $\mathbf{L}\cdot\mathbf{S}$, where $\mathbf{L} = (L_x, L_y, L_z)$ is the angular momentum operator vector, and $\mathbf{S} = \frac{1}{2}\boldsymbol{\sigma} = \frac{1}{2}(\sigma_x, \sigma_y, \sigma_z)$ is spin operator vector, with $\sigma_i$ the $i$'th Pauli matrix. It is noteworthy that $v_{ks}$ is non-diagonal in spin space due to the spin-orbit coupling term.

The interactions of electrons with the laser are described in the velocity gauge, where we employ the following vector potential:

$$\mathbf{A}(t) = f(t)\frac{cE_0}{\omega}\sin(\omega t)\hat{\mathbf{e}} \qquad (6)$$

where $f(t)$ is an envelope function (see the Appendix for details), $E_0$ is the field amplitude, $\omega$ is the carrier frequency, and $\hat{\mathbf{e}}$ is a unit vector that is generally elliptically polarized. Note that we neglect ion motion and assume the frozen nuclei approximation (i.e. omitting phononic excitations). This is expected to be a very good approximation in attosecond to femtosecond timescales, especially for heavy atoms. The KS equations of motion are solved in a real-space grid representation with Octopus code [60–62]. From the time-propagated KS states we calculate time-dependent observables of interest, including the total electronic current, $\mathbf{J}(t) = \frac{1}{\Omega}\int_\Omega d^3r\, \mathbf{j}(\mathbf{r},t)$, where $\Omega$ is unit cell volume and $\mathbf{j}(\mathbf{r},t)$ is the microscopic time-dependent current density:

$$\mathbf{j}(\mathbf{r},t) = \sum_{n,\mathbf{k},\alpha}\left[\varphi_{n,\mathbf{k},\alpha}^{KS\,*}(\mathbf{r},t)\left(\frac{1}{2}\left(-i\boldsymbol{\nabla}+\frac{\mathbf{A}(t)}{c}\right)+[V_{ion},\mathbf{r}]\right)\varphi_{n,\mathbf{k},\alpha}^{KS}(\mathbf{r},t) + c.c.\right] + \mathbf{j}_m(\mathbf{r},t) \qquad (7)$$

, where $\mathbf{j}_m(\mathbf{r},t)$ is the magnetization current density (which after spatial integration vanishes and does not contribute to $\mathbf{J}(t)$). $\mathbf{J}(t)$ is also used to obtain the HHG spectra, $\mathbf{I}(\Omega) = |\int dt\, \partial_t \mathbf{J}(t) e^{-i\omega t}|^2$. The spin expectation values are calculated as $\langle\mathbf{S}(t)\rangle = \langle\psi_{n,\mathbf{k}}^{KS}(t)|\mathbf{S}|\psi_{n,\mathbf{k}}^{KS}(t)\rangle$, and are used to track the spin dynamics in the system. All other technical details about the numerical procedures are delegated to the Appendix.

This numerical approach is employed in exemplary benchmark materials for exploring light-driven magnetization dynamics. The main example used throughout the text is the two-dimensional (2D) topological insulator, bismuthumane (BiH) [63]. BiH is comprised of a monolayer of bismuth atoms arranged in a honeycomb lattice (see illustration in Fig. 1(a)). The bismuth atoms are capped by hydrogen atoms that are covalently bonded to the bismuth $p_z$ orbitals in a staggered configuration that preserves inversion symmetry, but breaks some of the mirror planes of the honeycomb lattice. The electronic structure of BiH is strongly affected by spin-orbit coupling (SOC) – without SOC it exhibits Dirac cones in the K and K' high symmetry points, but SOC opens a large topological gap with nonzero Berry curvature throughout the Brillouin zone (see Fig. 1(b)) [63]. Each band is spin degenerate, and the ground-state is non-magnetic. BiH is an ideal candidate for exploring light-driven magnetic phenomena because the bismuth ions induce a relatively large SOC, and the system is an insulator which allows strong non-resonant nonlinear responses (the direct gap is 1.35 eV within the local spin density approximation). As we will show below, all of our results are independent of the topological insulator character of BiH.

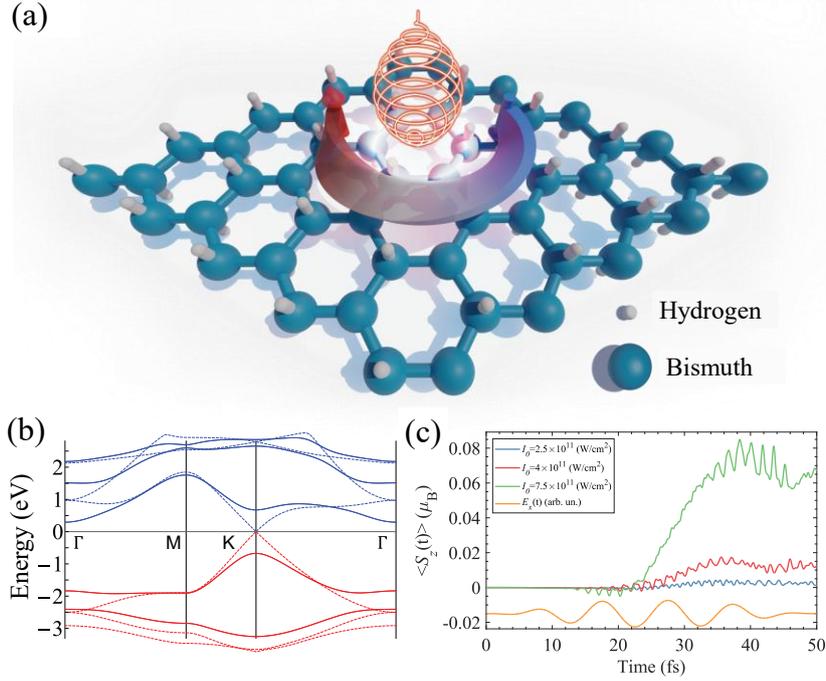

FIG. 1. Ultrafast turn-on of magnetism in monolayer BiH. (a) Illustration of the hexagonal BiH lattice and schematic illustration of the ultrafast turn-on of magnetism – an intense femtosecond laser pulse is irradiated onto the material, exciting electronic currents that through spin-orbit interactions induce magnetization and spin flipping. (b) Band structure of BiH with and w/o SOC (red and blue bands indicate occupied states and unoccupied states, respectively). In the SOC case, each band is spin degenerate. (c) Calculated spin expectation value, $<S_z(t)>$, driven by circularly polarized pulses for several driving intensities (for a wavelength of 3000nm). The $x$-component of the driving field is illustrated in arbitrary units to convey the different timescales in the dynamics.

## III. SUB-CYCLE TURN-ON OF MAGNETIZATION

We calculate the electronic response of BiH to intense circularly-polarized laser pulses (polarized in the monolayer $xy$ plane) with a carrier wavelength of 3000nm, and intensities in the ranges of $10^{11}$-$10^{12}$ W/cm$^2$ (see illustration in Fig. 1(a)). The corresponding carrier photon energy of 0.41 eV is well below the band gap, guaranteeing that the dominant light-matter response is non-resonant (at least four photons are required to excite an electron from the valence to the conduction band). These conditions result in HHG with harmonics up to the ~30$^{th}$ order corresponding to photon energies of ~12eV being emitted (see Appendix B). The circularly polarized drive imparts angular momenta onto the electronic system through the light-matter coupling term, and a combination of intraband acceleration and interband recollisions lead to the HHG emission [43,64]. It is noteworthy that due to a 6-fold improper-rotational symmetry in BiH, only harmonic orders of $6n \pm 1$ are emitted for integer $n$, which follows from dynamical symmetry selection rules [65]. As we will later show, similar selection rules can be derived for the total electronic excitation and the spin expectation values, which play a significant role in the magnetization dynamics.

The interesting question that we now explore is whether the laser-driven electronic angular momenta can be converted to a net magnetization. While such conversion was recently shown in resonant conditions [35], it is not clear if it can be obtained non-resonantly. Moreover, it is unknown whether this can be achieved in a system with fully spin-degenerate bands (precluding spin-selectivity *via* optical transitions between spin-split bands). Figure 1(c) shows the time dependent expectation value of spin along the *z*-axis, $\langle S_z(t) \rangle$, for several laser intensities. The BiH system is initially in a non-magnetic ground-state with $\langle \mathbf{S}(t=0) \rangle = 0$, but shows an onset of magnetization about 12 femtoseconds after it starts interacting with the laser. The characteristics of the magnetic response can be described by two main features: (i) sub-cycle fast oscillations of the spins, and (ii), a slower build-up of the magnetization that occurs over several laser cycles. By calculating the total occupation of electrons in just the up or down part of the spinors we verify that the induced magnetization indeed results from spin-flipping processes (see Appendix B). In other words, during the interaction with the laser spin-down electrons are flipped into a spin-up state. By reversing the helicity of the driving laser, one obtains the opposite picture with a conversion of up to down spins. We note that these transitions cannot occur from a resonant optical transition because the photon energies are well below the gap.

The very fast oscillations observed in Fig. 1(c) indicate that there are attosecond magnetization dynamics involved. For circularly polarized driving these attosecond spin-flipping processes accumulate over the laser cycle on a timescale of about ten femtoseconds, yielding a net magnetization. Notably, the response strongly depends on the driving power. Figure 1(c) shows that for stronger driving a larger net magnetization can be obtained (as high as ~$0.1\mu_B$). This result hints to the active mechanism at play: stronger driving increases the angular momentum of the excited electrons, which increases the contribution of the SOC term. This is also supported by calculations that show that the induced magnetic response diminishes with the driving ellipticity (see Fig. 2(a)). The process is thus analogous to the inverse Faraday effect, but in non-resonant and non-perturbative conditions. Figure 2(b) presents the scaling of the magnetic response with intensity – for weaker driving it follows a quartic dependence with the field amplitude, indicating a 4-photon response (the minimal number of photons required to excite an electron from the valence to the conduction band in these conditions), but above ~$4 \times 10^{11}$ W/cm$^2$ this dependence breaks down and behaves non-perturbatively. Interestingly, we note that the induced magnetization saturates for ellipticities between 0.2-0.4, i.e. the magnetic response stops increasing with ellipticity for that parameter range (Fig. 2(a)). This behavior differs from that of the inverse Faraday effect and reflects the extreme nonlinearity of the magnetization.

The onset time for the magnetization also strongly depends on the driving power and shows a non-perturbative nonlinear dependence (see Appendix B). Overall, this suggests that the spin-flipping processes are directly driven by the electronic excitations to the conduction band (because these are initiated by tunneling), which subsequently undergo additional laser-induced acceleration in the bands that lead to spin-flipping. We further support this picture with a *k*-space and band-resolved projection of the magnetization density, which validates that the regions around the K and K' points (minimal band gap positions) are the dominant contributors to the induced magnetization, and that the first valence and conduction bands have the largest contribution (see Appendix). We highlight that these observed dynamics differ from previously reported results in the perturbative resonant regime – the non-resonant driving here does not directly excite a spin-selective optical transition, and spin-spilt bands do not play any role.

Next we analyze the electronic excitations that allow for these spin flipping processes. Figures 2(c,d) plot the spectral components of the net magnetization of the system, i.e. the Fourier transform of $\langle S_z(t) \rangle$, *vs.* the driving intensity and wavelength. The main observation from Fig. 2(c) and (d) is that the spin excitation is inherently connected to the laser drive – remarkably, its spectral components are comprised of harmonics of the laser carrier frequency, and only $6n$ harmonics (for integer $n$) are allowed in the circularly polarized driving case. The Appendix presents results for elliptical driving which also support this conclusion, but where $2n$ (even) harmonics are allowed. This result is important for several reasons: (i) it directly proves that the laser-driven electron dynamics are converted to magnetization (*via* SOC), because otherwise the spin would not evolve in temporal harmonics of the laser. (ii) It allows to tune the temporal behavior of the spin-flipping by changing the driving parameters. (iii) It establishes the main microscopic mechanism for the ultrafast spin-flipping processes, which involves attosecond sub-cycle excitations of electrons from the valence to the conduction band. The excited electrons are subsequently driven by the laser field that imparts angular momentum, which is converted by the SOC term to spin flipping torques. In the Appendix we show that the total number of electrons excited to the conduction band over time evolves with the same exact symmetries as the spectral components of the magnetization (e.g. 6-fold in the circularly polarized driving case), and also derive the origin of this effect. Moreover, we confirm that the onset time for the magnetization dynamics is inherently connected to the electronic excitation. Thus, the two processes of strong-field tunneling between the bands, and spin flipping, are interlinked.

Notably, this mechanism follows the first steps in the HHG process in solids (i.e. tunneling of electrons to the conduction band, and subsequent acceleration in the bands) [43,64], but where spin and SOC play the additional role in driving a magnetic response. Overall, the onset of ultrafast magnetization is enabled by the fact that in the strong-field regime, electrons acquire relatively large momenta (i.e. large $\langle \mathbf{L} \rangle$), and that this happens repeatedly every laser cycle. The fact that electrons are driven in an extremely nonlinear manner that is inherently sensitive to attosecond sub-cycle timescales allows possibilities of atto-magnetism. Indeed, Fig. 2 shows that the electron's spin can oscillate with frequencies as high as the $42^{nd}$ harmonic of the laser, which promises incredible potential for atto-magnetism (that corresponds to ~100 attosecond magnetization dynamics, provided that the lower orders of the response can be suppressed). In the Appendix we validate that the induced magnetism is not driven by electronic correlations, which tend to slightly reduce the magnitude of the phenomena (as expected due to enhanced scattering). We note that since our simulations do not include sufficient dephasing and relaxation channels (because of the use of semi-local approximations to the XC functional, and because we do not incorporate electron-phonon couplings in the simulation), the electronic and spin excitations do not fully decay in our calculations. In realistic experimental conditions, we expect that these magnetic states will live for several tens of femtoseconds before decohering.

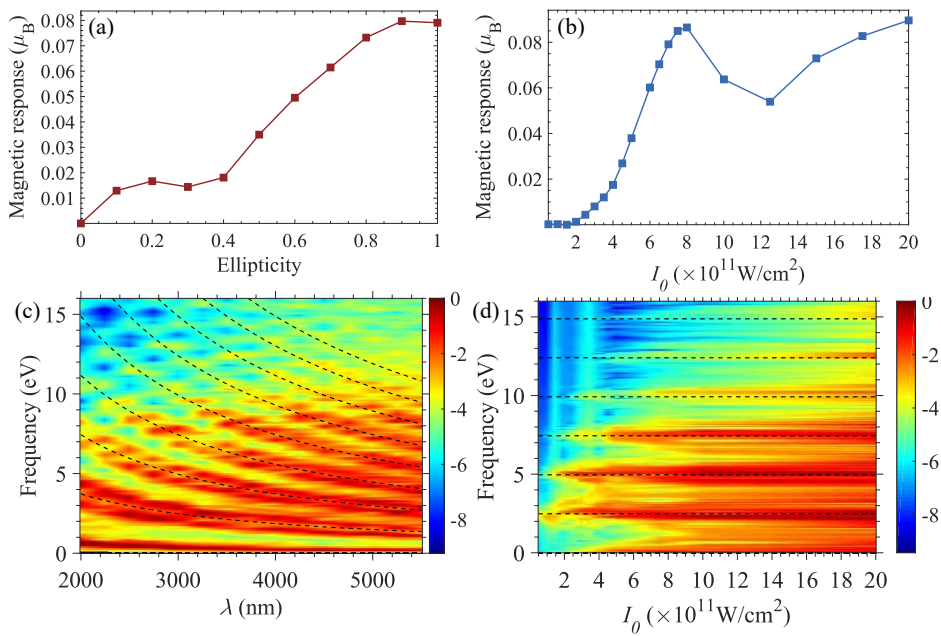

Fig. 2. Induced magnetic response for elliptical driving in BiH for changing laser parameters. (a) Maximal induced magnetization *vs.* the driving laser ellipticity, for driving power of $7\times10^{11}$ W/cm$^2$ and wavelength of 3000nm, where the elliptical major axis is transverse to the Bi-Bi bonds. (b) Same as (a) but for circularly polarized driving *vs.* laser power. (c) Spectral components of $<S_z(t)>$ *vs.* driving wavelength (calculated for an intensity of $5\times10^{11}$ W/cm$^2$, presented in log scale). Dashed black lines indicate $6n$ harmonics of the driving laser carrier frequency (for integer $n$), showing that the spin's temporal evolution is driven by the laser, and that its symmetries are connected to the light-driven electronic excitations. (d) Same as (c) but for changing laser intensity (for a driving wavelength of 3000nm). Higher intensities and longer wavelengths are shown to lead to higher frequency components, indicating a faster magnetic response.

## IV. ATTOSECOND MAGNETISM

We now further analyze the very fast oscillations of magnetization seen in Fig. 1(c). For circularly polarized driving, they are overlayed on top of the dominant slow response that builds up the net magnetization from cycle to cycle. That is, the dominant spin response is a zeroth order multipole. Consequently, it is very difficult to measure these fast oscillations experimentally, or to utilize them for applications. Still, the high energy spectral components in $\langle \mathbf{S}(t) \rangle$ appear be quite dominant compared to the weaker perturbative response, as long as there is a way to suppress the zeroth order slow contribution (see for instance Fig. 2(d)).

In order to try and extract this response we now explore linearly-polarized driving, where the in-plane polarization axis is given by the angle $\theta$, which is the offset angle from the *x*-axis (that is transverse to the Bi-Bi bonds). Importantly, in this case the cycle-averaged total angular momentum of the laser-matter system remains zero. In this respect, one expects no net magnetization to evolve, such that the zeroth order multipole of $\langle \mathbf{S}(t) \rangle$ should vanish. At the same time, intuition would dictate that no magnetization dynamics should occur at all, since even if one considers timescales shorter than a laser cycle the driving field is linearly-polarized and does not contain angular momenta. Nevertheless, Fig. 3(a) presents the temporal evolution of $\langle S_z(t) \rangle$ for several driving intensities, which shows a strong magnetic response that rapidly oscillates in time (on attosecond timescales). What is the origin of this transient magnetism?

Figure 3(b) presents the spectral components of the spin evolution *vs.* the in-plane laser polarization angle, which enables us to pinpoint the source of the effect. Clearly, the response follows fundamental symmetries of the material system – whenever the laser is polarized along a plane of BiH that exhibits a mirror or a 2-fold rotational symmetry the magnetic response fully vanishes. At this stage we recall that for these driving conditions, electrons are accelerated in the bands, which generates high harmonics and nonlinear currents in the system. The nonzero Berry curvature in BiH drives a transverse anomalous electric current [44,64,66–68]. However, from fundamental symmetries these transverse currents must vanish along the same high symmetry axes [65]. Figure 3(c) presents the HHG spectral components that are polarized only transversely to the main driving axis *vs.* the in-plane driving angle, which verifies this result. Notably, the emitted HHG radiation polarized transverse to the driving axis, and the spectral components of the magnetic response, are extremely similar (compare Figs. 3(b) to (c)). We thus conclude that a transverse current component is essential for generating the anomalous magnetic response. In accordance with the mechanism described above, this result is clear – the electronic system must acquire a nonzero angular momentum, $\langle \mathbf{L} \rangle$, since only then the SOC term can initiate a net magnetism (given that the system starts out in a non-magnetic state). Such angular momenta is only obtained if electrons are driven in at least two transverse axes. Thus, the anomalous currents and Berry curvature in this case act as a key ingredient for the magnetic response, since they initialize transverse electron motion that mimics the effects of circularly polarized driving. Indeed, if the SOC term is turned off the magnetic response completely vanishes (not shown).

We stress that after the laser matter interaction has ended, the system is left in a non-magnetic state as expected (the magnetization along each half cycle cancels out), removing the zeroth order spin response. Thus, linearly polarized driving enables the higher order spin dynamics to be uncovered, and even dominate over the slower perturbative responses. Arguably, one of the most exciting consequences is that the magnetic response oscillates very rapidly during the interaction with the laser. For instance, in an extreme case we observe the magnetization flipping from a local maximum of ~0.01$\mu_B$ to a local minimum of ~-0.01$\mu_B$ within just 411 attoseconds (see Fig. 3(a)).

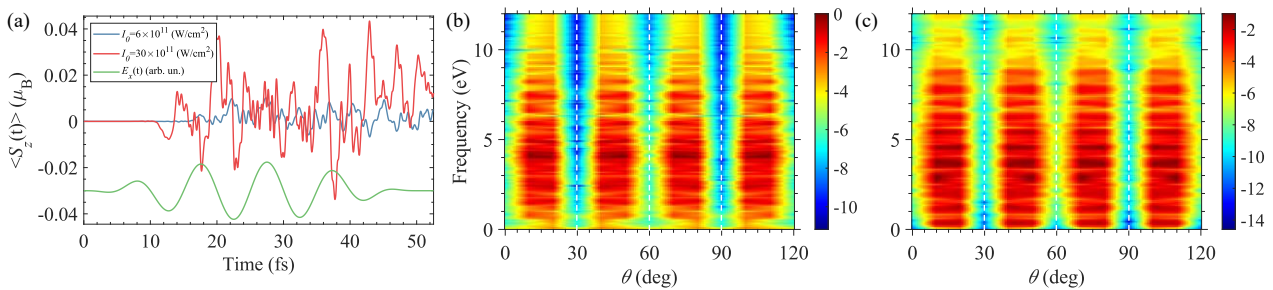

FIG. 3. Attosecond light-driven magnetic response in BiH. (a) Exemplary calculations of $\langle S_z(t) \rangle$ driven by intense linearly-polarized pulses for several driving intensities (for a driving wavelength of 3000nm, and a polarization angle of $\theta=10^0$), showing attosecond timescale magnetization dynamics. The driving field is illustrated in arbitrary units to convey the different timescales in the dynamics. (b) Spectral components of $\langle S_z(t) \rangle$ *vs.* driving angle. The magnetic response identically vanishes along high symmetry axes (indicated by white dashed lines). Plot calculated for a driving intensity of $2 \times 10^{11}$ W/cm$^2$ and a wavelength of 3000nm. (c) The HHG spectra polarized transversely to the driving laser axis, *vs.* the driving laser polarization axis in the monolayer plane. The transverse anomalous currents identically vanishe for the same driving conditions as the magnetic response, indicating that the two are connected. In (b) and (c) the spectral power is presented in log scale.

## V.  OTHER MATERIALS

At this point one may wonder about the generality of our results, since we studied light-driven dynamics in a 2D topological insulator. Thus, a legitimate question is whether the obtained ultrafast magnetization is a topological feature, and if it is also applicable in 3D bulk solids. To address these questions we explore the ultrafast magnetic response in MoTe$_2$, which is a transition metal dichalcogenide, and bulk Na$_3$Bi [69]. Specifically, we consider the 2H bulk phase and the H monolayer phases of MoTe$_2$, which are non-topological and have non-magnetic ground-states. Na$_3$Bi on the other hand has a topological nature, and was recently shown to be a Dirac semimetal with bulk Dirac cones appearing at finite momenta [69,70]. We consider it because it is a peculiar example for a material system with strong SOC, a non-magnetic ground-state, and inherently vanishing Berry curvature, but which still permits generation of transverse anomalous currents due to its hexagonal structure.

Figure 4(a) presents an exemplary temporal evolution of $\langle S_z(t) \rangle$ driven by an intense linearly-polarized pulse in a monolayer of H-MoTe$_2$. Figure 4(b) presents the spectral components of $\langle S_z(t) \rangle$ for linear driving *vs.* the in-plane driving angle. Clearly, similar ultrafast magnetic responses are obtained; albeit, they are slightly weaker because MoTe$_2$ exhibits weaker SOC. Thus, a main conclusion is that the attosecond-based magnetism is driven in non-magnetic materials independently on if the bands have nonzero Chern numbers or not. In general, very similar nonlinear behavior is obtained in MoTe$_2$ for all driving conditions (see Appendix C). Figure 4(c) presents the corresponding HHG spectra polarized transverse to the laser driving. Very good agreement between the two is obtained just as was seen in BiH (regardless of the different space group of MoTe$_2$ and BiH), confirming the generality of the results outlined above. In the Appendix we also present results for the bulk phase, which shows similar responses.

Similarly, in the Appendix we show that for linearly polarized driving there is a strong magnetic response in Na$_3$Bi. This is despite the fact that it has uniformly vanishing Berry curvature [69], and results from the hexagonal lattice that permits nonzero transverse currents to first order of the driving field (whereas Berry curvature is typically associated with a second-order nonlinear response), as long as it is not driven along high symmetry axes. Indeed, along mirror planes there are no transverse currents in Na$_3$Bi, and consequently, also no light-induced magnetization dynamics. Thus, our results indicate a general mechanism that enables attosecond magnetism in otherwise non-magnetic systems, including 3D bulk solids and 2D systems. The main ingredients for this response are: (i) highly non-resonant strong-field driving, (ii) strong spin-orbit coupling, and (iii) generation of anomalous transverse currents to the driving which are permitted by crystal symmetries, and which effectively yield a strong oscillating orbital angular momentum.

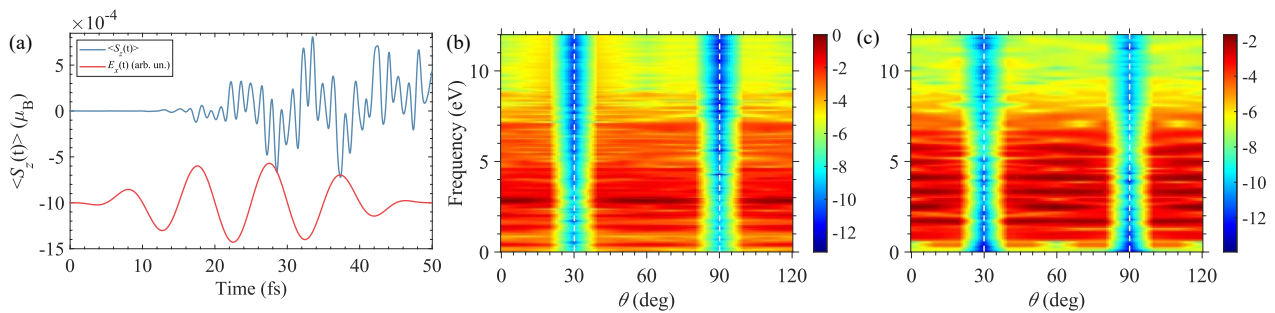

FIG. 4. Ultrafast turn-on of magnetization and attosecond magnetization dynamics in monolayer H-MoTe$_2$ (topologically trivial). (a) Exemplary calculation of $<S_z(t)>$ driven by a linearly-polarized pulse with an intensity of $2\times10^{11}$ W/cm$^2$, a wavelength of 3000nm, and polarized at $\theta=20^0$. The driving field is illustrated in arbitrary units to convey the different timescales in the dynamics. (b) Spectral components of $<S_z(t)>$ *vs.* driving angle in the same conditions as (a). (c) The HHG spectra polarized transversely to the driving laser axis, *vs.* the driving laser polarization axis in the monolayer plane. In (b) and (c) the spectral power is presented in log scale. Dashed white lines indicate high symmetry axes.

## VI. PUMP-PROBE CIRCULAR DICHROISM

Lastly, we present a potential experimental set-up capable of measuring these attosecond magnetic responses. This set-up is based on a pump-probe geometry, where the pump is an intense femtosecond pulse that excites magnetization dynamics (just as in Eq. (6)), and the probe is an attosecond extreme ultraviolet (XUV) pulse that is circularly (or elliptically) polarized. By measuring the time-resolved circular dichroism (CD) in the absorption (or transmission) spectra, one can detect attosecond magnetism (for numerical details see the methods section). Notably, since the ground-states for all of the material systems we examined are non-magnetic, the CD is zero if the system is not pumped. Thus, any nonzero signal immediately indicates the presence of magnetization, making the detection scheme potentially simpler and background-free.

Figure 5(a) presents an exemplary spectrum for BiH driven by a circularly-polarized pulse – the strong pump field induces changes to the imaginary part of the material's dielectric function. When this driven state of matter is probed with left circularly polarized (LCP) or right circularly polarized (RCP) pulses, there is a noticeable deviation in the response. The subtraction defines the CD, which can be further normalized to the ellipticity of the probe pulse in a given spectral region (because the probe pulse has a finite duration it is not perfectly circular). Depending on the conditions, the ellipticity-normalized CD can reach up to 50% changes of the dielectric function in equilibrium, which should be well within experimental detectability. Figure 5(b) presents the attosecond-resolved CD for the linearly-driven case. A strong CD signal oscillates rapidly in accordance with the induced magnetization (peak to minima changes occur on a timescale of ~500 attoseconds), illustrating that the attosecond magnetization dynamics should be experimentally accessible.

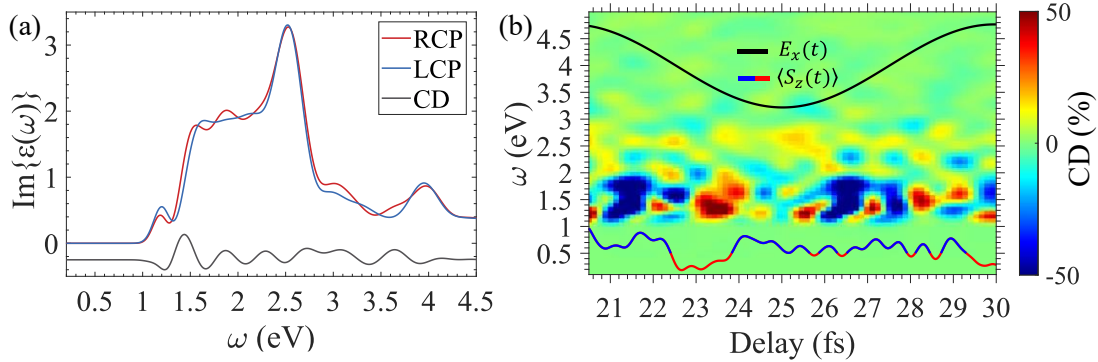

FIG. 5. Pump-probe time-resolved CD absorption spectroscopy in BiH. (a) Imaginary part of the dielectric function for the driven system (pumped with a circularly polarized pulse with a wavelength of 3000nm and intensity $7.5\times10^{11}$ W/cm$^2$), probed with either left (LCP) or right (RCP) helical probes. The driven system is temporally probed one cycle before the end of the pump laser field. The CD curve represents the subtraction between the RCP and LCP curves and is shifted down for clarity. Note that due to its finite duration, the probe pulse has a non-uniform ellipticity in this frequency region, which is not taken into account in (a). (b) Attosecond-resolved CD in the linear driving case (for a wavelength of 3000nm and intensity $3\times10^{12}$ W/cm$^2$, driven at an angle of $\theta=20^0$). The $x$-component of the driving electric field and the induced magnetization is plotted out of scale for reference, where positive/negative parts of the induced magnetization are highlighted by blue/red colors in correspondence with the CD. The CD in (b) is normalized by the ellipticity of the probe pulse for each frequency region.

## VII. SUMMARY AND OUTLOOK

To summarize, we investigated light-matter interactions between intense non-resonant femtosecond laser pulses and solids with strong spin-orbit coupling. Through state-of-the-art *ab-initio* calculations, we demonstrated that the attosecond timescale excitation and acceleration of electrons in the bands (which involves highly nonlinear multi-photon processes) is converted to magnetism by the SOC term. With circularly polarized driving, this induces a net magnetization that typically turns on within ~16 femtoseconds. Consequently, we establish a new regime of femto-magnetism where paramagnetic materials can be transiently transformed into magnetic states with non-resonant driving. Remarkably, even in linearly-polarized driving conditions there are significant magnetization dynamics during the interaction with the laser pulse, which are enabled by light-driven anomalous currents in materials that permit a transverse optical response. These dynamics evolve intrinsically on very ultrafast timescales (much faster that previously described) of ~500 attoseconds, as they result from the extreme nonlinear response of electrons to the driving lightwave itself, on a sub-cycle level. We studied the connection between the symmetries of the laser-matter system and the induced nonlinear magnetic response, showing that the speed of the magnetization dynamics can be tuned with the laser parameters. To our knowledge, these are the fastest known magnetization dynamics in solids, which are enabled by the extreme nonlinearity in the strong-field regime, and the unique linearly polarized drive that effectively removes slower terms in the magnetic response. They should pave the way towards attosecond control of magnetism, and motivate utilizing strong-field physics as an avenue for nonlinear spintronics with enhanced degrees of control over higher order spin and spin-photon interactions. Lastly, we showed that these phenomena should be experimentally detectable with pump-probe attosecond transient absorption experiments, utilizing circular dichroism [34,71–73].

It is worth discussing some possible applications and extensions of our results. First, while we employed here simple quasi-monochromatic laser pulses, our results are more general. In that respect, utilizing more complex waveforms such as bi-chromatic fields should enable enhanced control over magnetism. This simply follows from the enhanced control over electron dynamics that such fields offer [46,74,75]. This possibility is especially exciting because it could lead to even further increase of the speed of the magnetization dynamics, and to controlling ultrafast magnetization by tuning the laser phases (i.e. a form of coherent control). Second, by using few-cycle pulses we expect that one could induce a net magnetization even with linearly-polarized pulses, since then anomalous contributions from sequential half-cycles would not cancel out. This would establish a magnetic analogue to transient injection currents that have recently been measured [41,50,51]. Lastly, since the magnetization is driven in the same conditions that allow for high harmonic generation, it could enable high harmonic spectroscopy for probing magnetism, which has not been possible before [57]. Looking forward, we expect our results to motivate more experimental and theoretical work in the field.

## ACKNOWLDENGEMENTS

This work was supported by the Cluster of Excellence Advanced Imaging of Matter (AIM), Grupos Consolidados (IT1249-19), SFB925, "Light induced dynamics and control of correlated quantum systems" and has received funding from the European Union's Horizon 2020 research and innovation programme under the Marie Skłodowska-Curie grant agreement No 860553. The Flatiron Institute is a division of the Simons Foundation. O.N. gratefully acknowledges the generous support of a Schmidt Science Fellowship.

## APPENDIX A: TECHNICAL DETAILS

We report here on technical details for calculations presented in the main text. We start with details of the ground state DFT calculations that were used for obtaining the initial KS states. All DFT calculations were performed using Octopus code [60–62]. The KS equations were discretized on a Cartesian grid with the shape of the primitive lattice cells. Atomic geometries and lattice parameters were taken from ref. [63] for BiH ($a=b=5.53$Å, and a Bi-H distance of 1.82Å), taken as $a=b=3.55$Å for H-MoTe$_2$, as $a=b=3.56$Å, $c=15.35$Å for 2H-MoTe$_2$, and as $a=b=5.448$Å, $c=9.655$Å for Na$_3$Bi [69]. In all cases the space-group symmetric primitive unit cell was employed (with the hexagonal lattice vectors residing in the *xy* plane). For monolayer systems, the *z*-axis (transverse to the monolayer) was described using non-periodic boundaries with a length of 60 Bohr. The KS equations were solved to self-consistency with a tolerance $<10^{-7}$ Hartree, and the grid spacing used was 0.39 Bohr for BiH, and 0.36 Bohr for MoTe$_2$ and Na$_3$Bi. We employed a Γ-centered 24×24×1 *k*-grid for BiH, of 30×30×1 *k*-grid for H-MoTe$_2$, 24×24×8 *k*-grid for 2H-MoTe$_2$, and 28×28×15 *k*-grid for Na$_3$Bi. Deep core states were replaced by Hartwigsen-Goedecker-Hutter (HGH) norm-conserving pseudopotentials [59].

For the time-propagation of the main equations of motion we employed a time-step of 4.83 attoseconds. The propagator was represented by a Lanczos expansion and *k*-point symmetries were not assumed. In the time-dependent calculations of the monolayer systems, we employed absorbing boundaries through complex absorbing potentials (CAPs) along the aperiodic *z*-axis with a width of 15 Bohr [76] and a magnitude of 1 a.u. We calculated the total electronic excitation induced in the system in a time-resolved manner ($n_{ex}(t)$) by projecting the KS-Bloch states onto the ground state system:

$$n_{ex}(t) = N_e - \sum_{n,n'\in VB} \sum_{\mathbf{k}\in BZ} w_{\mathbf{k}} |\langle \psi_{n',\mathbf{k}}^{KS}(t=0) | \psi_{n,\mathbf{k}}^{KS}(t) \rangle|^2 \qquad (8)$$

where $N_e$ is the total number of active electrons in the unit cell, the projections are performed onto the valence bands of the ground state system (i.e. $n' \in VB$), and the summation is performed in the entire first Brillouin zone (BZ). $n_{ex}(t)$ gives a measure for the number of excited electrons during the light-driven dynamics.

The envelope function of the employed laser pulse, *f(t)* from Eq. (1), was taken to be of the following 'super-sine' form [77]:

$$f(t) = \left(sin\left(\pi \frac{t}{T_p}\right)\right)^{\left(\frac{\left|\pi\left(\frac{t}{T_p}-\frac{1}{2}\right)\right|}{w}\right)} \qquad (9)$$

where $w=0.75$, $T_p$ is the duration of the laser pulse which was taken to be $T_p=5T$ (~29.3 femtoseconds full-width-half-max (FWHM) for 3000nm light), where $T$ is a single cycle of the fundamental carrier frequency. This form is roughly equivalent to a super-gaussian pulse, but where the field starts and ends exactly at zero amplitude, which is numerically more convenient.

Calculations of transient absorption spectroscopy employed the real-time propagation approach detailed in ref. [78]. We employed short XUV pulses as probes that were comprised of a set of 8 stepwise jumps in the vector potential (each giving a Dirac Delta function peak in the time-domain electric field), which had a rotating polarization direction. Each step was polarized at 45 degrees, and was separated by 62.9 attoseconds, with respect to its previous, resembling an optical centrifuge [79]. The total temporal duration of the probe pulse is thus 503.1 attoseconds, and each peak had an intensity of $10^{10}$ W/cm$^2$ (which gives an intensity of ~$10^8$ W/cm$^2$ in the frequency region of interest of 1-5eV). To change the helicity of the probe pulses the direction of rotation of the centrifuge was rotated, and for normalization purposes the ellipticity of the probe was calculated using stokes parameters [80] in the frequency region of interest. This configuration allows calculating the CD in a wide frequency range with attosecond temporal resolution while avoiding performing may separate calculations with changing carrier wavelength (because the step-like nature of the probe pulse has an infinite frequency content).

By calculating the light-driven current in the system we extracted the optical conductivity *via*:

$$\sigma_{ii}(\omega) = \frac{\tilde{J}_{probe,i}(\omega)}{\tilde{E}_{probe,i}(\omega)} \quad (10)$$

where $\tilde{J}_{probe,i}(\omega)$ is the Fourier transform of the total current in the system that is induced solely by the probe pulse. That is, in the time-domain $\mathbf{J}_{probe}(t)$ is defined as the subtraction of the total current that is calculated with the presence of the probe pulse, and the current that is calculated without a probe pulse that is driven solely by the pump. Here $\tilde{\mathbf{E}}_{probe}$ is the Fourier transform of the electric field vector of the XUV probe pulse, and $i,j$ are Cartesian indices. From the optical conductivity we extracted the dielectric function:

$$\varepsilon_{ii}(\omega) = 1 + \frac{4\pi}{\omega} \sigma_{ii}(\omega) \quad (11)$$

and the average dielectric function $\varepsilon(\omega) = \left(\varepsilon_{xx}(\omega) + \varepsilon_{yy}(\omega)\right)/2$. The CD was calculated between the imaginary part of the dielectric functions for a right-circular probe and a left-circular probe:

$$CD(\omega) = \text{Im}\{\varepsilon^+(\omega) - \varepsilon^-(\omega)\} \quad (12)$$

where +/- refers to left or right circularly polarized probes. Eq. (12) was evaluated for different pump-probe delays by changing the onset time of the probe pulse, where for the temporally-resolved plot in Fig. 5(b) we used steps of 500 attoseconds in the pump-probe delay grid and results were interpolated by splines on a denser grid. We also filtered the induced probe current, $\mathbf{J}_{probe}(t)$, with an exponential mask in the time domain to avoid numerical issues with the finite time propagation, and filtered $\varepsilon_{ii}(\omega)$ with an exponential mask below the band gap to remove issues of division by zero (because the probe pulse has zero spectral components at $\omega = 0$).

## APPENDIX B: ADDITIONAL RESULTS IN BiH

We present here additional complimentary results to those presented in the main text. Figure 6 presents the HHG spectra emitted from BiH in the same driving conditions as those that induce the transient magnetization discussed in the main text. Figure 6(a) presents the HHG spectra for circularly polarized driving while tuning the laser wavelength, showing that only $6n\pm1$ harmonics (for integer *n*) are emitted due to dynamical symmetry considerations [65]. Figure 6(b) presents the emitted spectra for elliptical driving, where the elliptical major axis is rotated within the monolayer plane. As seen, odd-only harmonics are emitted due to similar symmetry considerations [65]. Figure 6(c) presents the emitted HHG spectra *vs.* the driving ellipticity (changing from circular to linear) where the main elliptical axis is along the *x*-axis (transverse to the Bi-Bi bonds). In this case we can track the harmonic emission as the system transitions from the odd-only inversion symmetry to the 6-fold rotational symmetry obtained for circularly polarized driving. The same symmetries that constrain the HHG spectral components were shown in the main text to constrain the temporal evolution of the electronic and spin excitations, clarifying that the atto-magnetic response is directly driven by the laser.

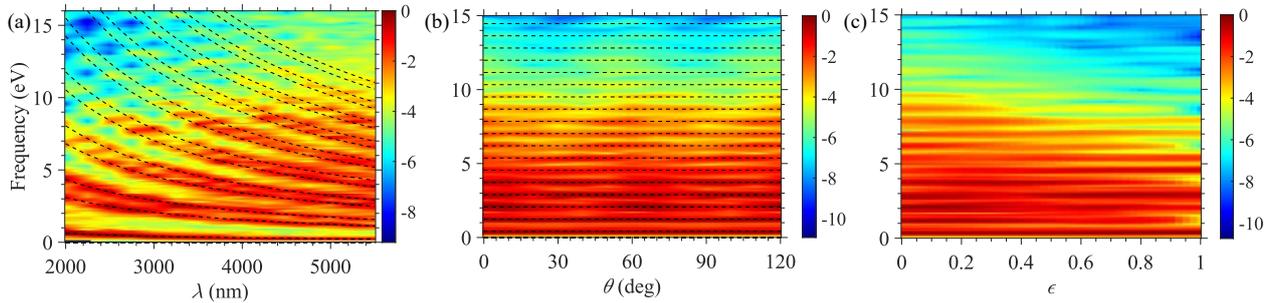

FIG. 6. HHG spectra and selection rules in BiH. (a) HHG spectra for circularly polarized driving *vs.* driving wavelength, calculated for $5\times10^{12}$ W/cm$^2$. Dashed black lines indicate $6n\pm1$ harmonics (for integer $n$). (b) HHG spectra for elliptical driving with an ellipticity of 0.2, *vs.* driving angle of the elliptical major axis in the monolayer plane (for a driving wavelength of 3000nm and intensity of $2\times10^{11}$ W/cm$^2$). Dashed black lines indicate odd harmonics. (c) HHG spectra for elliptical driving *vs.* the driving ellipticity, where the elliptical major axis is set at $\theta$=0, for similar laser parameters as (b). All spectra are presented in log scale.

Next, we further explore the femto-magnetic response in BiH. Figure 7(a) presents the difference in total occupations of spin-up and spin-down electrons from the ground state as the system evolves in time when driven by intense circular pulses (which is compensated for the small average ionization in both channels), $\Delta n_\alpha$. As is clearly shown, as the system builds up a magnetic response, the occupations of spin-down states is converted to up spins, or vice versa. In fact, the occupation of spin-up states *vs.* time is formally equivalent to the calculation of $\langle S_z(t)\rangle$. Thus, this result supports the mechanism responsible for this behavior, which involves a SOC-driven flipping of spins. Figure 7(b) presents a similar result but for circularly polarized driving with an opposite helicity, showing that there is perfect spin-helicity symmetry – exchanging light's helicity flips the magnetic response.

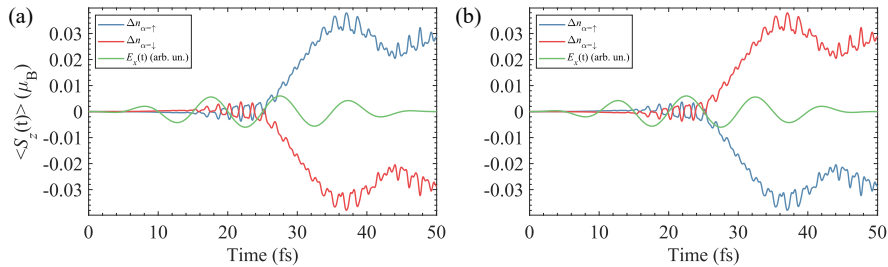

FIG. 7. Ultrafast magnetization dynamics in BiH. (a) Time-dependent occupations of spin-up and spin-down channels for circularly polarized driving (with a wavelength of 3000nm and intensity of $10^{12}$ W/cm$^2$). (b) same as (a) but for opposite driving light helicity. In all plots the *x*-component of the driving laser field is illustrated in arbitrary units for clarity.

We now present additional results that explore the nonlinear nature of the induced magnetism. Figure 8(a) presents the magnetization onset time (defined as the time for which the induced magnetization temporal derivative, $\partial_t\langle S_z(t)\rangle$, surpasses $10^{-6}\mu_B$ per atomic unit of time) *vs.* the driving intensity for the circularly polarized driving case. The onset time behaves highly nonlinearly with the pump power, which is additional corroboration for the nonlinear nature of the effect. Figure 8(b) presents the induced net magnetization in the same conditions as in Fig. 1(c) in the main text, with the induced total electronic excitation, $n_{ex}$. The plot connects the onset time of the magnetization with the light-induced excitations to the conduction band, as the two curves have a similar onset behavior. Moreover, the rapid attosecond dynamics become most prominent when the conduction band is highly populated. Similar results are obtained for other driving conditions.

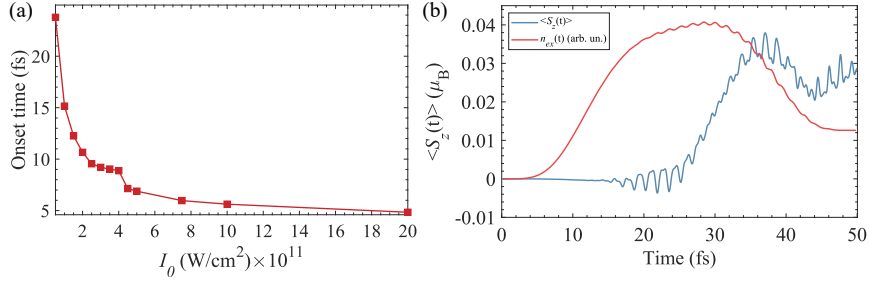

FIG. 8. Magnetization onset times and highly nonlinear nature of the magnetization dynamics in BiH. (a) Magnetization onset times *vs.* driving intensity for circularly polarized driving with a wavelength of 3000nm. (b) <$S_z$(t)> for circularly polarized driving with a driving wavelength of 3000nm and intensity of $5\times10^{11}$ W/cm$^2$, and total electronic excitation in the same scale.

We next explore the connected symmetries of the electronic excitation and the spin dynamics, which further support the conclusions presented in the main text. Figure 9(a) presents the spectral components of the electronic excitation, i.e. the Fourier transform of $n_{ex}(t)$ to the frequency-domain. This analysis allows tracking the temporal dynamics of the electronic excitation and seeing if it complies to similar symmetries as the induced magnetization. As seen in Fig. 9(a), for circularly polarized driving $n_{ex}(t)$ is comprised of only $6n$ harmonics of the driving laser (for integer $n$). This is a fundamental constraint that arises from dynamical symmetries [65], and complements the symmetries of the emitted HHG radiation (following $6n \pm 1$ selection rules, see Fig. 6). Essentially, Eq. (8) evaluates the sum of projections of the time-dependent wave functions onto the ground-state wave functions. However, because the dynamics of the wave functions uphold the improper-rotational dynamical symmetry in the light-matter system (which complies to $S_6 H\left(t + \frac{T}{6}\right) S_6^\dagger = H(t)$, where $S_6$ is a 6-fold improper rotation in the monolayer plane), this means that the occupations uphold $n_{ex}(t) \approx n_{ex}(t + T/6)$, where $T$ is the laser period, which leads to the $6n$ harmonic structure. This equation is correct up to a constant shift at zero frequency that has to do with non-periodic tunneling dynamics, and symmetry breaking due to the short duration of the laser pulse (which broadens the $6n$ harmonic peaks). The inherent reason that the selection rules for $n_{ex}(t)$ are different than for the HHG emission, is that $n_{ex}(t)$ is calculated with a parity-even projection operator, whereas the dipole operator that evaluates the HHG emission is parity-odd. Nonetheless, the two selection rules have similar origins. A completely identical selection rule is obtained for the expectation value of the total magnetization, $\langle S_z(t) \rangle$ (e.g. as seen in Fig. 2 in the main text, and in Fig. 9(b)), also allowing only $6n$ harmonics. It means that the two quantities of $n_{ex}(t)$ and $\langle S_z(t) \rangle$ are inherently connected, because the induced magnetization is physically driven by the excitations to the conduction band.

Figure 9(c) further presents the spectral components of $n_{ex}(t)$ for a system driven by an elliptically-polarized pulse. In this case, the light-matter system follows a dynamical inversion symmetry ($iH\left(t + \frac{T}{2}\right)i^\dagger = H(t)$, where i is an inversion operator), which means that the projections follow $n_{ex}(t) \approx n_{ex}(t + T/2)$, leading to even-only harmonics. This complements the odd-only HHG emission (see Fig. 6) and has an identical origin. Figure 9(d) presents the spectral components of $\langle S_z(t) \rangle$ for the same conditions, also showing even-only harmonics further supporting the generality of the mechanism presented in the main text. We obtained similar results that connect the symmetries of the light-driven currents, the electronic excitation, and the induced magnetization, in all explored conditions.

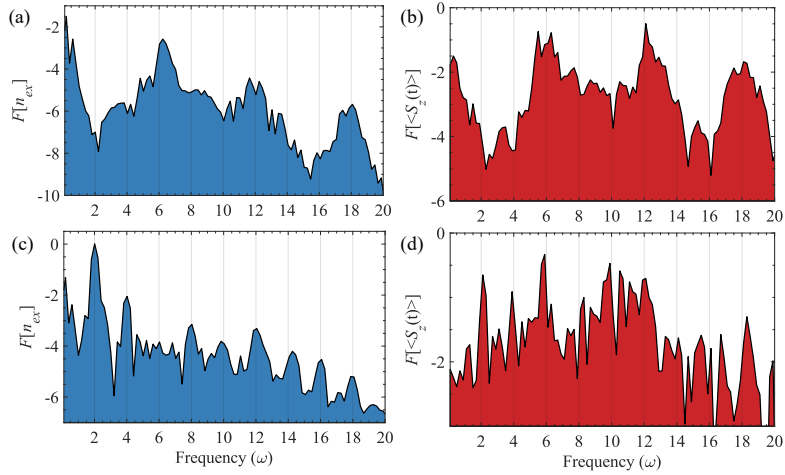

FIG. 9. Symmetries of the electronic excitation and their connection to magnetization dynamics in BiH. (a) Spectral components for $n_{ex}(t)$ plotted in arbitrary units in log scale for circularly polarized driving (with wavelength of 3000nm and intensity of $5\times10^{11}$ W/cm$^2$). The *x*-axis is given in units of the laser frequency, and even harmonics are indicated with grey lines. Only $6n$ harmonics are observed with circularly polarized driving (for integer $n$). (b) Same as (a) but for $<S_z(t)>$ in similar conditions, showing similar symmetry-based selection rules. (c) Same as (a) but for elliptical driving with an ellipticity of 0.5, a major elliptical axis angle of $\theta=30^0$, and driving intensity of $2\times10^{11}$ W/cm$^2$. Only even harmonics of the drive are observed. Same as (c), but for $<S_z(t)>$ in similar conditions, showing similar symmetry-based selection rules.

We now present the band- and *k*-space-resolved induced magnetization. The band-resolved contributions to the magnetization were calculated by projecting the time-dependent wave functions onto the ground-state wave functions, just as was done for $n_{ex}(t)$. However, in this case the projections were not summed in *k*-space and over the bands, and instead, the *k*-resolved projections for pairs of spin-degenerate bands were summed together after weighting the occupations by the expectation value $\langle S_z \rangle$ at that particular band and *k*-point. This gives a measure for the different contributions of each band and *k*-point to the induced magnetization, and for instance at *t=0* this analysis leads to identically zero magnetization in all bands and *k*-points (because the initial state is non-magnetic). Figures 10(a-d) presents these results for a circularly polarized driving case after the laser pulse ends for the first and second valence and conduction bands (in the notation where the 'first' bands includes two spin-degenerate bands, and so on). As is seen, the dominant contribution to the magnetization arises from regions near the minimal band gap (K and K' points). Moreover, the first valence and conduction band contribute stronger magnetic responses than other bands. This trend continues in higher bands (not presented). These results thus support that the first step in the mechanism behind the induced magnetization is electronic excitation to the conduction band, as discussed in the main text. In contrast to these results, Figs. 10(e,f) present the complementary *k*- and band-resolved magnetization after the interaction with a linear driving pulse for the first valence and conduction bands. Here, each region around K and K' contributes to both positive and negative magnetization, and the magnetization at K and K' is inverted, such that the net magnetism overages out to zero, as expected for linear driving.

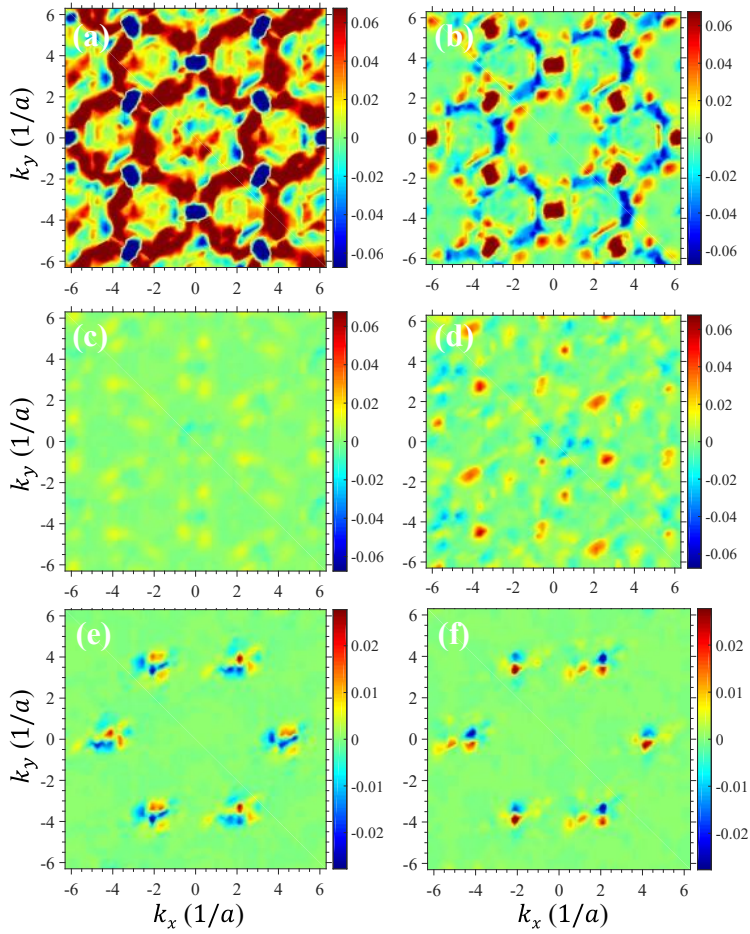

FIG. 10. $k$- and band-resolved projections of the light-induced net magnetization after the pulse. (a-d) Magnetization after circularly polarized driving with a wavelength of 3000nm and intensity of $7.5\times10^{11}$ W/cm$^2$. (a) First valence band, (b) first conduction band, (c) second valence band, (d) second conduction band. (e,f) Same as (a,b), respectively, but for linear driving with an intensity of $2\times10^{11}$ W/cm$^2$ and a polarization axis at $\theta=20^0$.

We also tested the role of correlations in the ultrafast induced magnetization dynamics. We have thus far employed TDSDFT in the local spin density approximation for the XC functional, which allows the *e-e* interactions to evolve dynamically in time. However, it is helpful to explore the role of correlations by employing the independent particle approximation (IPA), which is equivalent to freezing the time-evolution of the XC potential, and the time-evolution of the Hartree term. Within this approach the time propagation of all the KS-Bloch states is fully independent of each other, and dynamical *e-e* interactions are not included in the simulation. Figure 11 presents the spin expectation value in circularly polarized driving, comparing the IPA to the full TDSDFT calculation. As is clearly seen, the magnetization dynamics are not driven by correlations as a very similar response is obtained within the IPA.

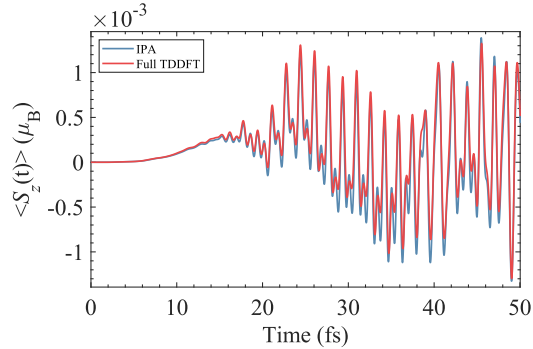

FIG. 11. $<S_z(t)>$ in BiH calculated with full TDSDFT compared to IPA for circularly polarized driving, with a wavelength of 3000nm and a intensity of $2\times10^{11}$ W/cm$^2$.

## APPENDIX C: ADDITIONAL RESULTS IN MoTe$_2$

We present here additional complimentary results to those presented in the main text for the MoTe$_2$ material system, both in monolayer and bulk phases. For the monolayer, Fig. 12(a) presents the spin expectation value driven by a circular pump, showing a femtosecond turn-on of magnetism that is in correspondence with the results obtained in the main text. The electronic excitation matches in onset time with the onset of magnetization (see Fig. 12(a)). Figure 12(b) presents the spectral components of $n_{ex}(t)$ in the same driving conditions, showing that in this case only $3n$ components (for integer $n$) are allowed (because a monolayer of H-MoTe$_2$ is 3-fold symmetric instead of 6-fold). Similarly, Fig. 12(c) presents the spectral components of $\langle S_z(t) \rangle$ that also support only $3n$ harmonics. These results arise from the different symmetries of MoTe$_2$ compared to BiH. It demonstrates the generality of our analysis.

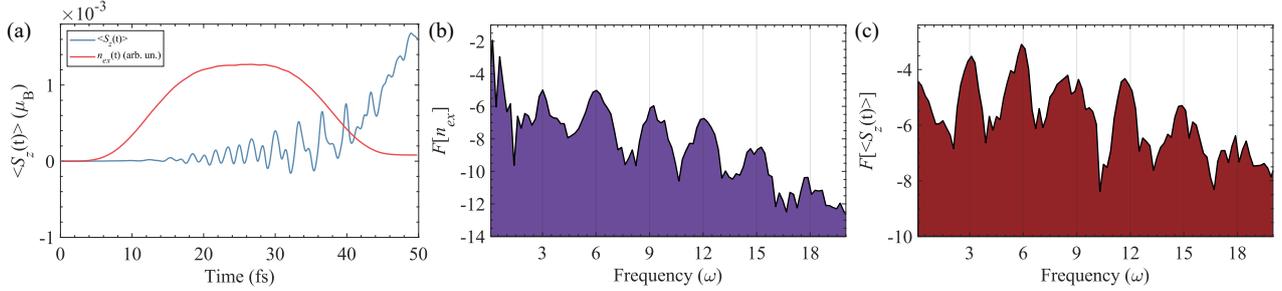

FIG. 12. Additional results for light-induced magnetization dynamics in monolayer H-MoTe$_2$. (a) $<S_z(t)>$ and $n_{ex}(t)$ for circularly polarized driving with a wavelength of 3000nm and intensity of $2\times10^{11}$ W/cm$^2$. (b) Spectral components of $n_{ex}(t)$ plotted in log scale for the same driving conditions as in (a). The $x$-axis is given in units of the laser frequency, and every 3$^{rd}$ harmonic is indicated with grey lines. Only $3n$ harmonics are observed with circularly polarized driving (for integer $n$). (c) Same as (b) but for $<S_z(t)>$ in similar conditions, showing similar symmetry-based selection rules. (b) and (c) are plotted in arbitrary units in log scale.

Figure 13 presents results for the 2H bulk phase of MoTe$_2$ for circular and linear driving. For the circular case (Fig. 13(a,b)) we obtain a femtosecond turn-on of the magnetic response which is comprised of $6n$ harmonics of the pump laser, owing to the inherent symmetries of the bulk 2H phase driven by circular light (the bulk phase exhibits a 6-fold rotation coupled to a glide symmetry along the $c$-axis). For the linear case we obtain attosecond magnetization dynamics, in similar spirit to the results in the main text, which comprise of even-only harmonics of the laser (see Fig. 13(c,d)). Thus, the results in the bulk phase support the results obtained in the monolayer systems, and demonstrate that the effect is valid in 3D bulk systems.

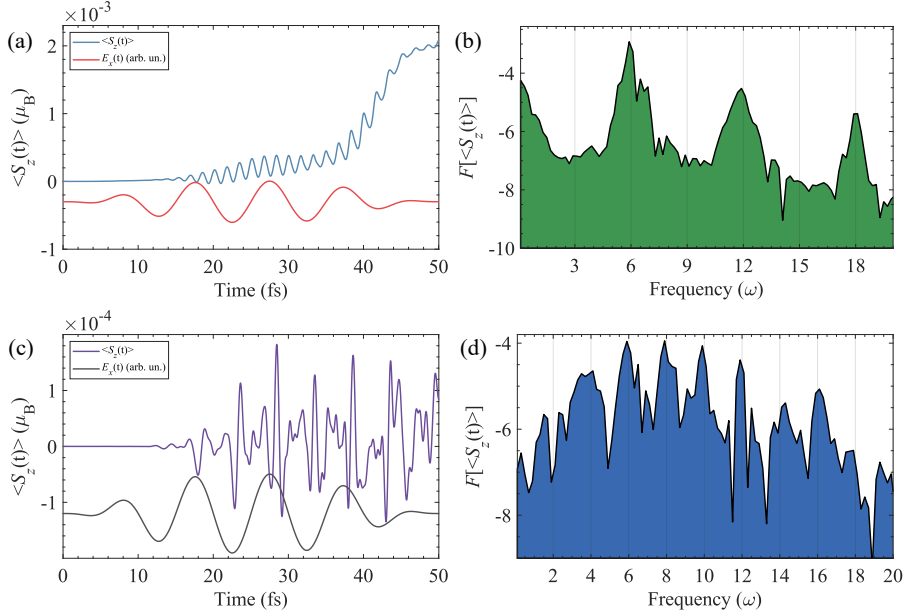

FIG. 13. Femtosecond turn-on of magnetization and attosecond magnetization dynamics in bulk 3D 2H-MoTe2. (a) $<S_z(t)>$ for circularly polarized driving with a wavelength of 3000nm and intensity of $2\times10^{11}$ W/cm$^2$. The *x*-component of the driving field is illustrated in arbitrary units to convey the different timescales in the dynamics. (b) Spectral components of $<S_z(t)>$ plotted in log scale for the same driving conditions as in (a). The *x*-axis is given in units of the laser frequency, and every 3rd harmonic is indicated with grey lines. Only 6*n* harmonics are observed with circularly polarized driving (for integer *n*). (c) Same as (a) but for linearly polarized driving with $\theta=10^0$. (d) same as in (b) but for the driving conditions in (c). Every 2nd harmonic is illustrated by grey lines, and only even harmonics are observed. (b) and (d) are plotted in arbitrary units in log scale.

## APPENDIX D: ADDITIONAL RESULTS IN Na$_3$Bi

We present additional complimentary results to those presented in the main text for Na$_3$Bi, the Dirac semimetal. Figure 14 shows the induced magnetization dynamics driven by intense linearly-polarized light in the hexagonal planes for two different driving angles (either along a high-symmetry axis, or not). As clearly shown, when Na$_3$Bi is not driven along high symmetry axis it permits a strong oscillating magnetization. We verified that this is a result of nonzero transverse currents that permit transient orbital angular momentum (not presented). On the other hand, along high symmetry axes the transverse currents are not permitted, leading to an identically vanishing magnetism. Thus, these results support the generality of our conclusions, even in materials with vanishing Berry curvature.

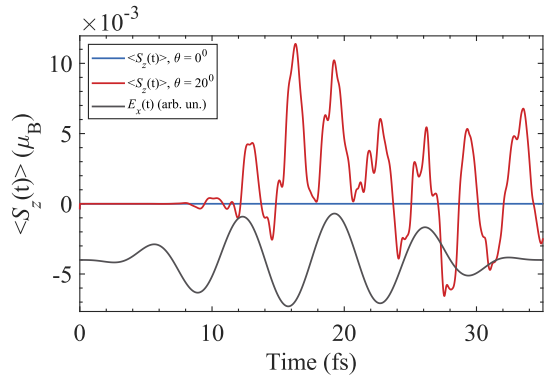

FIG. 14. Attosecond magnetization dynamics in the Dirac semimetal Na$_3$Bi for linearly polarized driving with a wavelength of 2100nm and intensity of $7\times10^{11}$ W/cm$^2$. The *x*-component of the driving field is illustrated in arbitrary units to convey the different timescales in the dynamics. The plots show the magnetic response for driving in the *xy* plane either along a mirror axis ($\theta=0$), or at an angle of $20^0$ from that axis.